%% file: FSQ-18-001_temp.tex
\pdfoutput=1

\documentclass[11pt,twoside,a4paper,cmspaper,final,collab]{cms-tdr}
\def\svnVersion{f6cc152-D}\def\svnDate{2019/10/04}\def\cmsCernNoTag{CERN-EP-2019-146}\def\cmsCernDate{\today}\def\cmsMessage{"Published in the European Physical Journal C as \href{http://dx.doi.org/10.1140/epjc/s10052-019-7402-3}{\doi{10.1140/epjc/s10052-019-7402-3}.}"}

\begin{document}\cmsNoteHeader{FSQ-18-001}

\hyphenation{had-ron-i-za-tion}
\hyphenation{cal-or-i-me-ter}
\hyphenation{de-vices}
\RCS$HeadURL$
\RCS$Id$

\newlength\cmsTabSkip\setlength{\cmsTabSkip}{1ex}

\newlength\cmsFigWidth
\ifthenelse{\boolean{cms@external}}{\setlength\cmsFigWidth{0.85\columnwidth}}{\setlength\cmsFigWidth{0.4\textwidth}}
\ifthenelse{\boolean{cms@external}}{\providecommand{\cmsLeft}{upper\xspace}}{\providecommand{\cmsLeft}{left\xspace}}
\ifthenelse{\boolean{cms@external}}{\providecommand{\cmsRight}{lower\xspace}}{\providecommand{\cmsRight}{right\xspace}}
\cmsNoteHeader{FSQ-18-001}

\title{Measurement of the average very forward energy as a function of the track multiplicity at central pseudorapidities in proton-proton collisions at $\sqrt{s}=13\TeV$}

\titlerunning{Very forward energy as a function of the track multiplicity at central pseudorapidities}

\newcommand{\CASTORAcceptance} {{$-6.6<\eta<-5.2$}}
\newcommand{\epos}{\textsc{epos~lhc}}
\newcommand{\sibyll}{\textsc{sibyll}}
\newcommand{\qgsjetII}{\textsc{QGSJetII}}

\date{\today}

\abstract{
The average total energy as well as its hadronic and electromagnetic
components are measured with the CMS detector at pseudorapidities
$-6.6<\eta<-5.2$ in proton-proton collisions at a centre-of-mass energy
$\sqrt{s}=13\TeV$. The results are presented as a function of the charged
particle multiplicity in the region $\abs{\eta}<2$. This measurement is
sensitive to correlations induced by the underlying event structure
over a very wide pseudorapidity region. The predictions of Monte Carlo
event generators commonly used in collider experiments and ultra-high energy
cosmic ray physics are compared to the data.
All generators considered overestimate the fraction of energy
going into hadrons.
}

\hypersetup{%
pdfauthor={CMS Collaboration},%
pdftitle={Measurement of the average very forward energy as a function of the track multiplicity at central rapidities in proton-proton collisions at sqrt(s)=13 TeV},%
pdfsubject={CMS},%
pdfkeywords={CMS, CERN LHC, forward phase space, CASTOR, underlying event}
}

\maketitle

\section{Introduction}
\label{sec:intro}

{\tolerance=450
The description of inclusive hadron production in high energy
hadron-hadron collisions remains subject to significant theoretical
uncertainties.  At TeV energies the dominant source of secondary
particle production is the fragmentation of quarks and gluons in
semihard scattering~\cite{PhysRevD.40.1446}, referred to
as minijet production. However, various processes that cannot
be directly calculated from first principles in quantum chromodynamics
(QCD) also contribute to particle production, \ie multiparton
interactions (MPIs), and fragmentation of the remnants. Together with
initial- and final-state radiation these additional particle production
mechanisms are typically referred to as the underlying event and
are modelled phenomenologically in Monte Carlo (MC) event generators
with parameters tuned using
data~\cite{Sjostrand:1987su,Corke:2010yf,Skands:2014pea}. In addition,
especially in the forward phase space, diffractive processes play an
important role~\cite{N.Cartiglia:2015gve}.  Furthermore, final-state
parton rescattering effects, a possible hydrodynamical phase transition, or
other collective phenomena can impact and modify particle production
in hadron-hadron collisions at high energies~\cite{PhysRevC.89.064903}.\par}

The energy carried by particles emitted into the very forward region (\CASTORAcceptance)
covered by the CASTOR calorimeter~\cite{Andreev:2010zzb} of the CMS
experiment was shown to be a powerful probe of the activity of
the underlying event~\cite{Chatrchyan:2013gfi,Sirunyan:2017nsj}.
For the first time measurements presented in this paper correlate the
hadronic energy at very forward rapidities to the central region in
proton-proton collisions, offering a new approach to the study of
hadron production at the CERN LHC.
Such measurements over a very large rapidity interval provide additional
information on the underlying event compared to those based only  on the
central region, \eg Refs.~\cite{Aaboud:2017fwp,Sirunyan:2017vio}.

The very forward region covered by the data contains the
highest energy densities, $\rd{}E$/$\rd\eta$~\cite{Sirunyan:2018jwn,Aaij:2012pda}, so far observed in proton-proton collisions at the LHC. Therefore, the present results can
improve event generators used in simulations of extensive air showers induced by cosmic
rays at ultra-high
energies~\cite{dEnterria:2011twh}. Specifically, current air shower
simulations are known to significantly underestimate muon
production~(see Ref.~\cite{Dembinski:2019uta} and references therein).
The fraction of the energy going into the production of electrons or photons rather than
long-lived hadrons has a crucial impact on the muon
production rate in extensive air showers, see Ref.~\cite{Ulrich:2010rg}.
Since CASTOR consists of separate electromagnetic and hadron
calorimeters, the data presented here
provide new information that may improve understanding of muon production
in air showers.

\section{Experimental setup and Monte Carlo simulation}
\label{sec:setup}
The main feature of the CMS apparatus is a superconducting solenoid
of 6\unit{m} internal diameter that can provide a nominal magnetic field of
3.8\unit{T}.  Within the solenoid volume in the central region are a
silicon pixel and strip tracker, a lead tungstate crystal
electromagnetic calorimeter, and a brass and scintillator hadron
calorimeter. Muons are measured in gas-ionisation detectors embedded
in the steel return yoke. The central detectors of CMS are
complemented by calorimeters in the forward direction, which all rely
on the detection of Cherenkov photons produced when charged particles
pass through their active quartz components. The ``hadron forward''
(HF) calorimeters cover the pseudorapidity interval
$3.0<\abs{\eta}<5.2$ and use quartz fibres embedded in a steel
absorber. The CASTOR calorimeter is a sampling calorimeter composed of
layers of fused silica quartz plates and tungsten absorbers. It is
located on only  one side of CMS and covers the
region \CASTORAcceptance. CASTOR is segmented into 16 azimuthal
towers, each with 14 longitudinal channels. The two front channels
have a combined depth of 20 radiation lengths and form the
electromagnetic section of each tower. The remaining 12 channels
constitute the hadronic section. The full depth of a tower amounts to
10 hadronic interaction lengths. A more detailed description of the
CMS detector, together with a definition of the coordinate system used
and all relevant kinematic variables, can be found in
Ref.~\cite{Chatrchyan:2008zzk}. A detailed description of the CASTOR
calorimeter is given in
Refs.~\cite{Andreev:2010zzb,Gottlicher:2010zz,Sirunyan:2017nsj}. For
triggering purposes, the Beam Pickup Timing for the eXperiment (BPTX) devices
were used~\cite{Khachatryan:2016bia}.

{\tolerance=1150
The data are compared to a broad range of model predictions covering different
parameter tunes as well as entirely different physics approaches. The models considered
are \PYTHIA8~\cite{Sjostrand:2014zea} (version 8.212) with tune
CUETP8M1~\cite{Khachatryan:2015pea}, and tune 4C~\cite{Corke:2010yf},
combined with the MBR~\cite{Ciesielski:2012mc} model to describe
diffractive processes. The data are also compared to the predictions
of \epos~\cite{Pierog:2013ria} and \sibyll~2.1~\cite{Ahn:2009wx}. For these models, a
detailed Monte Carlo simulation of the CMS detector response is
performed with the \GEANTfour~\cite{Agostinelli2003250} toolkit. The
simulated events are processed and reconstructed in the same way as
the collision data. Furthermore, predictions
by \qgsjetII.04~\cite{PhysRevD.83.014018}, \sibyll~2.3c~\cite{Riehn:2017mfm},
\PYTHIA8 tune CP5~\cite{GEN-17-001},
and \HERWIG~7.1~\cite{Bahr:2008pv,Bellm:2015jjp} with the default tune
for soft interactions~\cite{Gieseke:2016fpz} are also compared to the
data. These simulations are produced only at the generator level. A
forward folding method is developed to compare generator-level
simulations to the data. This technique can be used to compare any
model or theoretical prediction to the data and will
be described in detail.\par}

\section{Data analysis and systematic uncertainties}
\label{sec:analysis}
This analysis is based on data recorded during the
low-luminosity startup operation of the LHC in June 2015, at a
proton-proton centre-of-mass energy of 13\TeV. In this period the CMS
solenoid was turned off.  The data correspond to an
integrated luminosity of 0.22\nbinv, with an average
proton-proton interaction probability of about 30\% per bunch
crossing.

The event selection criteria are optimised to select inelastic
collision events with minimal bias.
The residual contribution of electronic noise and beam background in these events is well below 1\%.
Events were selected online with an unbiased trigger
requiring only the presence of two colliding bunches.
The offline event selection requires activity in the HF calorimeters: at least
one tower with reconstructed energy larger than 5\GeV in either the positive or
negative HF calorimeter. In addition, at
least one reconstructed track with $\abs{\eta}<2$ is required in the CMS pixel detector.
A modified tracking algorithm from Ref.~\cite{Khachatryan:2015jna} is used in the
absence of a magnetic field. Information from the pixel detector
is used to reconstruct straight tracks.  Signals in all three
layers of the pixel detector are required to lie within a cone of
radius $R = \sqrt{\smash[b]{(\Delta\phi)^2+(\Delta\eta)^2}}
= 0.02$ (where $\phi$ is the azimuthal angle in radians) around the
reconstructed track. The efficiency to find more than two hits in the
pixel detector drops quickly for $\abs{\eta}>2$; the search for tracks
is therefore limited to $\abs{\eta}<2$. Tracks are retained if they
originate from the expected interaction region and are linked to at least one
interaction vertex. This pixel track reconstruction has an efficiency of about 76\% and
a probability of $\approx$5\% of spurious tracks for charged particles
with a transverse momentum \pt larger than 200\MeV.

To reject events with more than one simultaneous proton-proton
interaction (pileup), an additional constraint on the reconstructed interaction
vertices is applied. Events with two reconstructed vertices are
rejected if the vertices are separated by more than 0.5\cm along the $z$ axis.
This minimises the rejection of events with
high particle multiplicity, where the reconstruction may create
multiple spurious vertices.  The probabilities for events to have
additional collisions is evaluated in both data and simulation
to be 1.5\% (visible vertex) and 2.3\% (invisible vertex).
The correction of these background events is not
straightforward, since the correction depends on the track
multiplicity in the central region as well as on the model used in
simulation. Therefore, the contribution from pileup events to the forward energy
is considered part of the systematic uncertainty of the
measurement.

The total energy deposited in CASTOR is obtained by summing the
energy measured in each calorimeter tower above the noise
threshold, which is determined independently for each tower
and varies between 2 and 2.5\GeV. On average, 76\% of the showers due to single electrons
or photons are contained within the electromagnetic section of CASTOR, and
single hadrons are 71\% contained in the hadronic section.
Moreover, for a given particle energy, the energies deposited by hadron-induced
showers are smaller than electron-induced showers, which is known as noncompensation.
These properties were precisely measured with a test beam and are implemented
in the detector simulation. It was previously shown that the energy deposited
in the corresponding sections of CASTOR can serve as good estimators for the
particle-level energy of electrons/photons and hadrons~\cite{Sirunyan:2017nsj}.
The electromagnetic and hadronic energies of a given event are defined as the energies
deposited in the corresponding detector sections of CASTOR, and the total energy as the sum of both.

The events are classified according to the number of
reconstructed charged tracks from the vertex.
The average total, electromagnetic, and hadronic energy
per event is calculated for each track multiplicity bin. The present data
make it possible to study track multiplicities up to
150. The statistical uncertainties of the energy measurement are
below 2\%, much smaller than the systematic
uncertainties. The most
important sources of systematic uncertainties are described in the following and
 are summarised in Table~\ref{tab:Systematics}:

\textit{CASTOR energy scale.} The energy scale uncertainty of CASTOR
is 17\%~\cite{Sirunyan:2017nsj}. The energy scale is determined using
a calibration procedure based on SPS test-beam data, LHC beam halo
muon events, a cross-calibration to the HF calorimeters, and LED test
pulses, in combination with a precise detector alignment.  The
precision is currently limited by systematic effects related to the
modelling and understanding of particle shower cascades in the
calorimeter ranging from GeV to TeV energies. \\
\textit{CASTOR intercalibration.} The relative intercalibration
is performed using the measured response of each channel to single LHC
beam halo muon events, which were recorded with a dedicated trigger during LHC
interfill periods. This procedure is limited by the available muon
statistics. For a measurement of the total energy, the uncertainty caused by intercalibration is averaged over the whole
calorimeter and is 2--3\%. For the determination of the
electromagnetic and hadronic energy fractions, on the other hand, the effect of relative
calibration becomes more significant. Dedicated studies
based on full detector simulations of collision events
demonstrate that
the  observed average shape of the longitudinal shower absorption in the calorimeter is consistent with only
a slight overestimation of electromagnetic energies, and a corresponding
underestimation of hadronic energies.
We determine a maximum decrease of the electromagnetic energy by 8\%
and a corresponding increase of the hadronic energy by 15\%, which are
included as systematic uncertainties.\\
\textit{Pileup rejection.} The uncertainty arising from the pileup contribution is
estimated by considering alternative vertex multiplicity selections;
events with exactly one reconstructed vertex, as well as events with
two vertices separated by less than 0.7\cm, are selected. These
changes mainly affect the high-multiplicity region and lead to a
systematic energy uncertainty of up to 10\% for multiplicity
$>$140. Collisions that do not create visible vertices in the detector
introduce an additional uncertainty that is below 0.8\%.\\
\textit{HF energy scale.} The uncertainty in the reconstructed HF energies
is 10\%~\cite{Chatrchyan:2011wm}. Varying the threshold for the event
selection from 5.0\GeV per HF calorimeter tower to 4.5 and 5.5\GeV
changes the average energy observed in CASTOR by less than 0.5\%.\\
\textit{Tracking.} The track reconstruction uncertainty has been
previously determined from studies comparing data and
simulation~\cite{Khachatryan:2015jna}. The uncertainties in the
tracking and vertexing efficiencies affect the number of reconstructed
tracks by 1.8 and 2--3\%, respectively. These are combined linearly,
yielding a 5\% systematic uncertainty in the number of reconstructed
tracks. The effect in the average energy is below 5\%.

Most of the uncertainties described here are uncorrelated and are
therefore added in quadrature.
Moreover, in the measured ratios between electromagnetic and hadronic
energies the absolute energy scale uncertainty cancels, while the
intercalibration uncertainty introduces a particular anticorrelated
effect since a systematic decrease of the electromagnetic energy
causes an increase of the hadronic energy and vice versa.

\begin{table*}[tb]
\centering
\caption{Uncertainties in the average energies measured with the CASTOR calorimeter at the detector level. Ranges indicate the variation as a function of the track multiplicity.}
\label{tab:Systematics}
\begin{tabular}{lccc}
\hline
Source & Total energy & Electromagnetic energy & Hadronic energy \\
\hline
CASTOR energy scale & 17\% & 17\% & 17\% \\
CASTOR intercalibration & 2--3\% & $-$8\% & $+$15\% \\
HF energy scale & $<$0.5\% & $<$0.5\% & $<$0.5\% \\
Track reconstruction & 1--5\% & 1--5\% & 1--5\% \\
Pileup rejection & 1--8\% & 1--8\% & 1--10\% \\
Statistical uncertainty & 0.05--1.6\% & 0.06--1.9\% & 0.06--1.8\% \\[\cmsTabSkip]
Total & 18--19\% & 18--20\% & 20--26\% \\
\end{tabular}
\end{table*}

\section{Forward folding of model predictions}
\label{sec:folding}

The measured track multiplicity is distorted with respect to the
true charged particle multiplicity
by the effects of acceptance and
efficiency of the CMS pixel tracker. Likewise, the energies observed
in CASTOR are affected by the energy resolution and the response of
the calorimeter. In the present paper, the data are not corrected for these
effects, and should thus be compared to the results of a
full Monte Carlo detector simulation to compare with other
experimental data and to future model predictions.
For this purpose, a ``forward folding" approach is used here, in which
all known detector effects are applied to a given model prediction or theoretical
calculation. The forward folding approach
is chosen since it yields better systematic uncertainties
compared to an unfolding of these data.

At the generator level, events are selected that match the
detector-level event selection.  At least one charged particle with
$\pt>200\MeV$ is required within $\abs{\eta}< 2$. Furthermore, a
fractional momentum loss of the scattered proton of $\xi > 10^{-6}$ is
required. To determine $\xi$ all stable ($c\tau>1\cm$) final-state
particles are divided into two systems, X and Y, based on their
position with respect to the largest rapidity gap in the event. All
particles on the negative side of the largest gap are assigned to
system X, while the particles on the positive side are assigned to
system Y.  Based on this, we determine
$\xi=\max\left(M_{\mathrm{X}}^2/s,\,M_{\mathrm{Y}}^2/s\right)$, where
$M_{\mathrm{X}}$ and $M_{\mathrm{Y}}$ are the invariant masses of the
two systems.  The selection based on $\xi$ is relevant at very low
particle multiplicities, and leads to an optimal agreement with
the event selection as implemented at the detector level. It is also
consistent with previous CMS publications, \eg
Ref.~\cite{Sirunyan:2017nsj,Sirunyan:2018nqx}.

Four-dimensional migration matrices $k$ describing the probability
to reconstruct an event with central multiplicity
$N_{\mathrm{tracks}}$ and forward energy $E_{\mathrm{reco}}$ for given
values $N_{\mathrm{ch}}$ and $E_{\mathrm{true}}$ are calculated based on all
available Monte Carlo samples with full detector simulation.
At the
generator level, the central multiplicity $N_{\mathrm{ch}}$ is defined
as the number of stable charged final-state particles with
$\pt>200$\MeV and $\abs{\eta}< 2$, and the forward energy
$E_{\mathrm{true}}$ is defined as the sum of the energies of all
particles within \CASTORAcceptance{} except for muons and
neutrinos. At the detector level, the number of reconstructed tracks
with $\abs{\eta}<2$ is $N_{\mathrm{tracks}}$ and the reconstructed
energy in CASTOR is $E_{\mathrm{reco}}$.
The four-dimensional matrices $k_{ij}^{lm}$ are constructed with 20 bins in
$N_{\mathrm{ch}}$ and $N_{\mathrm{tracks}}$ ranging from 1 to 200
(dimensions $i$ and $l$)
, as well as 46 bins in $E_{\mathrm{true}}$ and $E_{\mathrm{reco}}$ ranging
from 0 to 10\TeV
(dimensions $j$ and $m$). The bin intervals
used at detector and generator level are identical. The range of $k$ is larger than that
used for the final results in order to allow for the effects of bin migration.
Final results are presented for $N_{\mathrm{tracks}}$ between
1 and 150.

All four components of $k$ have one extra underflow bin to
handle the event selection efficiency.
If an event does not pass the event selection criteria at the
generator level ($N_{\mathrm{ch}}\geq1$ and $\xi>10^{-6}$), it is
recorded in the underflow region with $N_{\mathrm{ch}}=0$ and
$E_{\mathrm{true}}=-1\GeV$. If an event is not selected at the
detector level (one HF tower above 5\GeV and
$N_{\mathrm{tracks}}\geq1$), it is recorded in the underflow region
with $N_{\mathrm{tracks}}=0$ and $E_{\mathrm{reco}}=-1\GeV$.  In this
way, the effects of inefficiencies and migrations from outside the
visible phase space are included in $k$.
For example, the selection efficiency for events having a specific
$N_{\text{ch}}$ and $E_{\text{true}}$ is the ratio of the number
of events without the underflow bin to the number of events with the
underflow bin.

Two-dimensional distributions,
$N_{\text{reco}}^{ij}$,
describing the event yields in bins $(i,j)$ of $N_{\mathrm{tracks}}$ and $E_{\mathrm{reco}}$
can then be obtained for any given event generator
or theoretical prediction by means of the following matrix multiplication:
\begin{equation}
N_{\text{reco}}^{ij} = \sum_{l,m} k_{ij}^{lm}  N_{\text{true}}^{lm},
\end{equation}
where
$N_{\text{true}}^{lm}$ is
the distribution of generator-level events in bins $(l,m)$ of
$N_{\text{ch}}$ and $E_{\text{true}}$. The average energy in each
track multiplicity bin is calculated from $N_{\text{reco}}^{ij}$
excluding the underflow bins,
and is compared to the data directly at the detector level. The
results obtained by using the forward folding method coincide with
those obtained with the full detector simulation to better than 1\%.

The matrix $k$ has a slight dependence on the $\eta$, \pt and multiplicity distributions
of the final-state particles in the
event generator used in the full detector simulation. To quantify this dependence, four
matrices are provided based on \PYTHIA8 tune CUETP8M1, \PYTHIA8
tune \textsc{4C+MBR}, \epos{}, and \sibyll~2.1. A fifth matrix is
obtained by averaging the matrices of these models and serves as the
central value for all forward-folded results. The spread of the
results obtained with the individual matrices is  an estimate
of the systematic uncertainty related to the model dependence; it is
mostly well below 5\%, but reaches 15\% in a few bins.
All five variations of $k$ are available in a
\textsc{rivet}~\cite{Buckley:2010ar} plugin. This way, the forward folding can be
applied to any other model prediction. Moreover, the full point-to-point
correlation of the model-related uncertainty can be studied.

\section{Results}
\label{sec:results}

\begin{figure*}[ptb]
\centering
\includegraphics[width=0.45\linewidth]{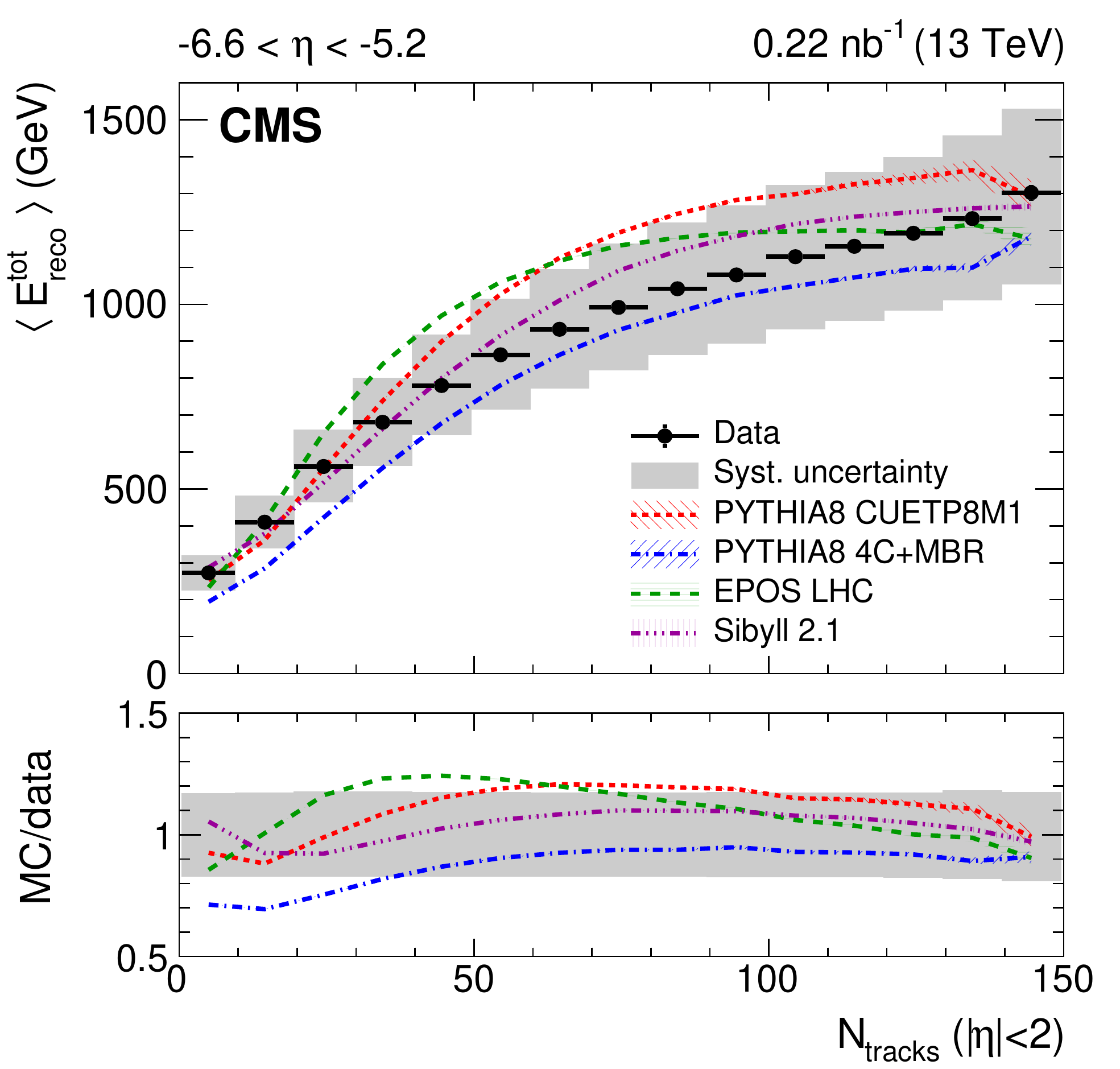}~
\includegraphics[width=0.45\linewidth]{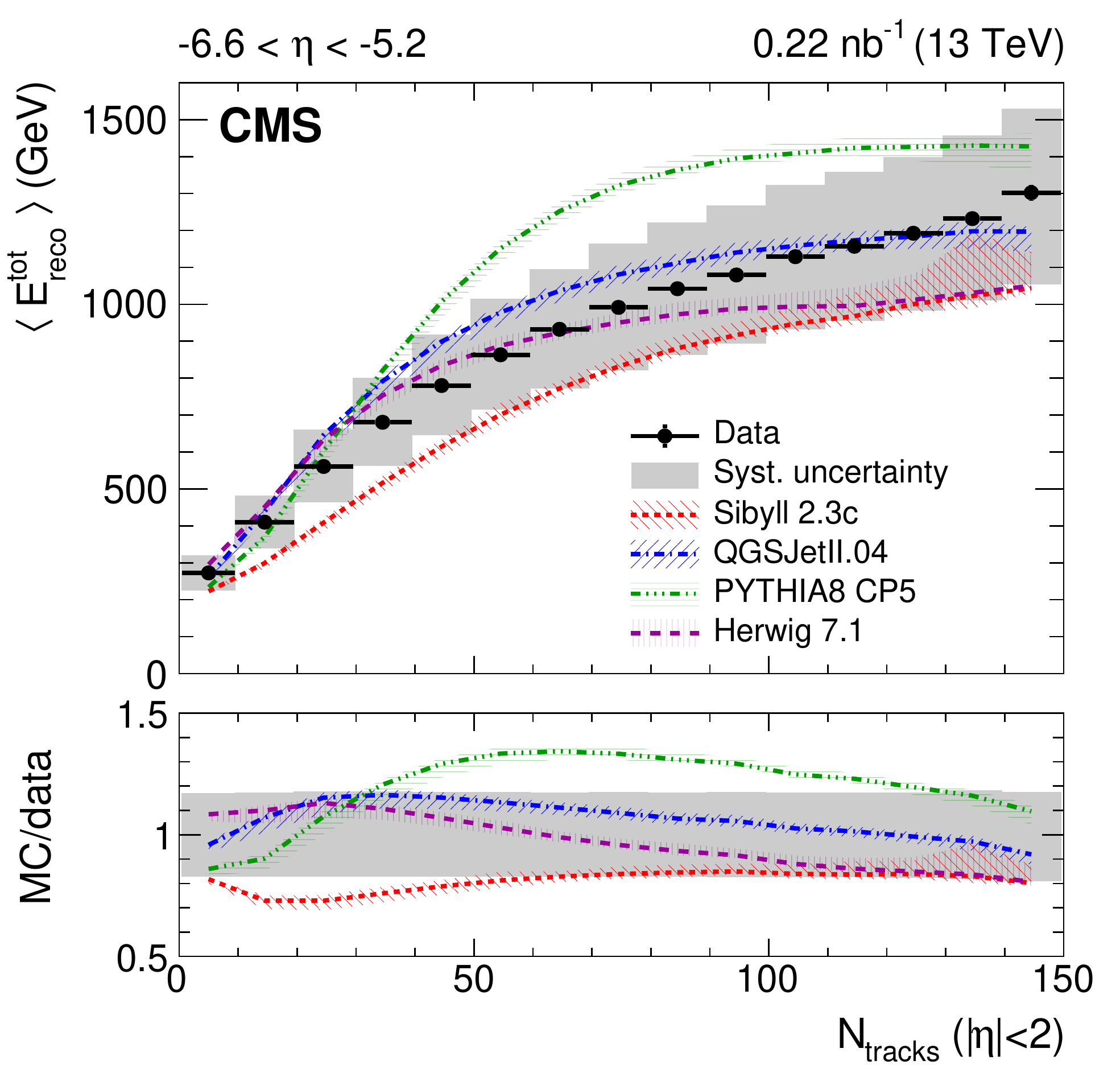}\\
\includegraphics[width=0.45\linewidth]{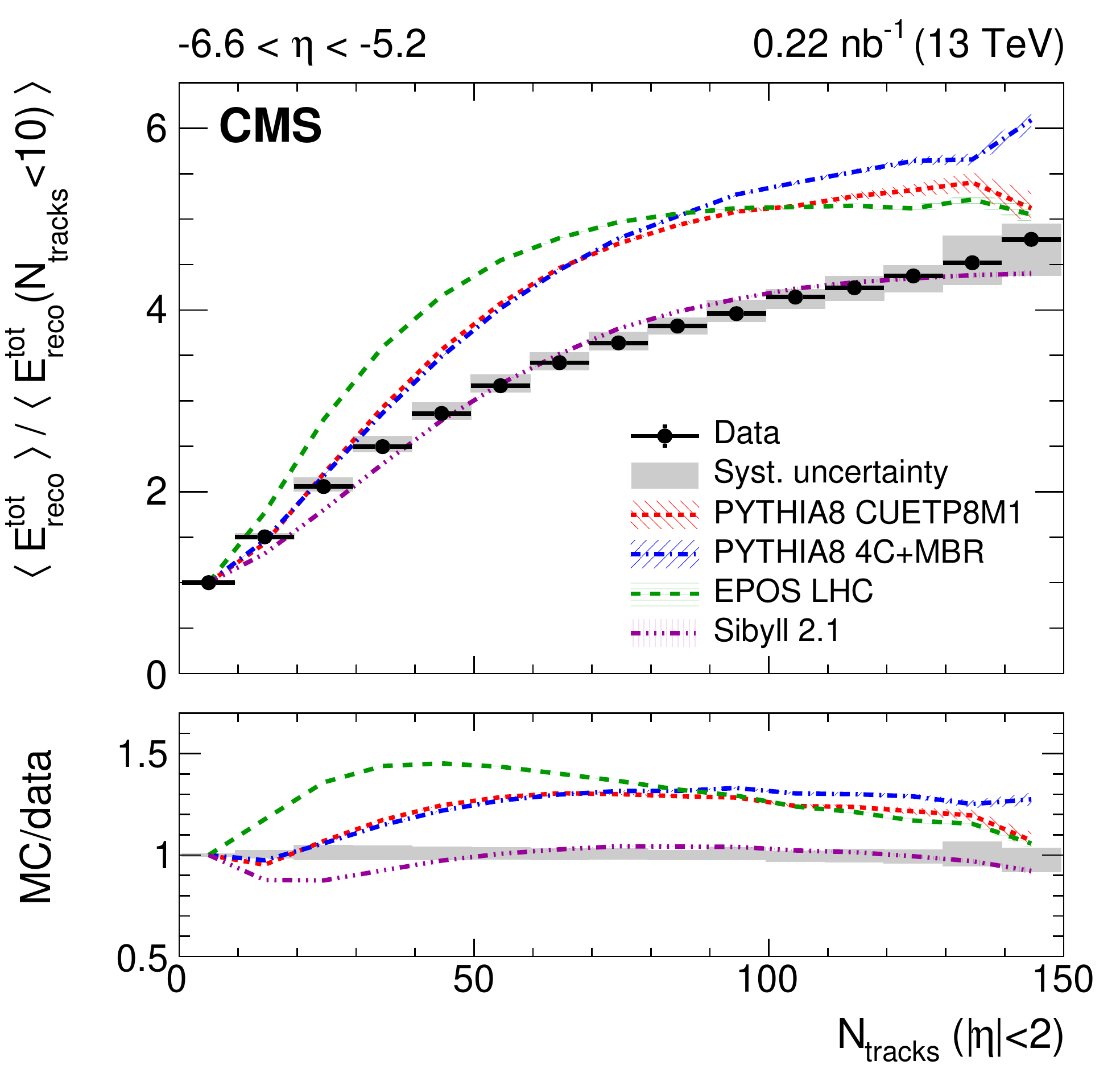}~
\includegraphics[width=0.45\linewidth]{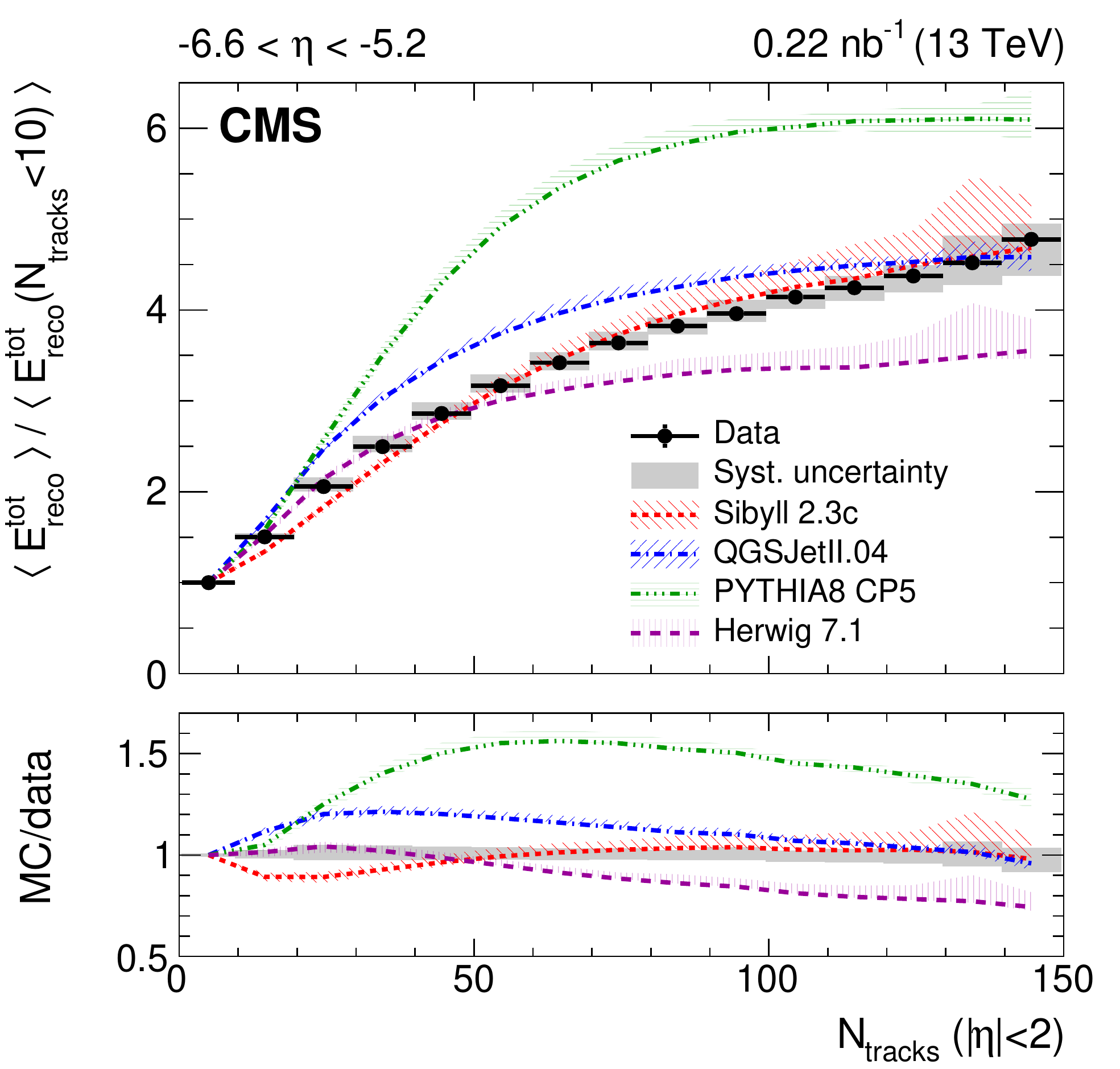}\\

\caption{Top panel: Average total energy reconstructed in the CASTOR calorimeter as a
function of the number of reconstructed tracks for $\abs{\eta}<2$.
Bottom panel: Average total energy reconstructed in the CASTOR calorimeter
normalised to that in the first bin ($N_{\mathrm{ch}}<10$) as a function
of the number of reconstructed tracks for $\abs{\eta}<2$. In all figures, the data
are shown as black circles and the corresponding systematic uncertainties
with a gray band; horizontal bars are used to indicate the bin width.
The predictions of various event generators are compared
to the data, which are the same in both panels. The bands associated with the
model predictions illustrate the model uncertainty.}
\label{fig:MFlowDetector_all}
\end{figure*}

\begin{figure*}[ptb]
\centering
\includegraphics[width=0.45\linewidth]{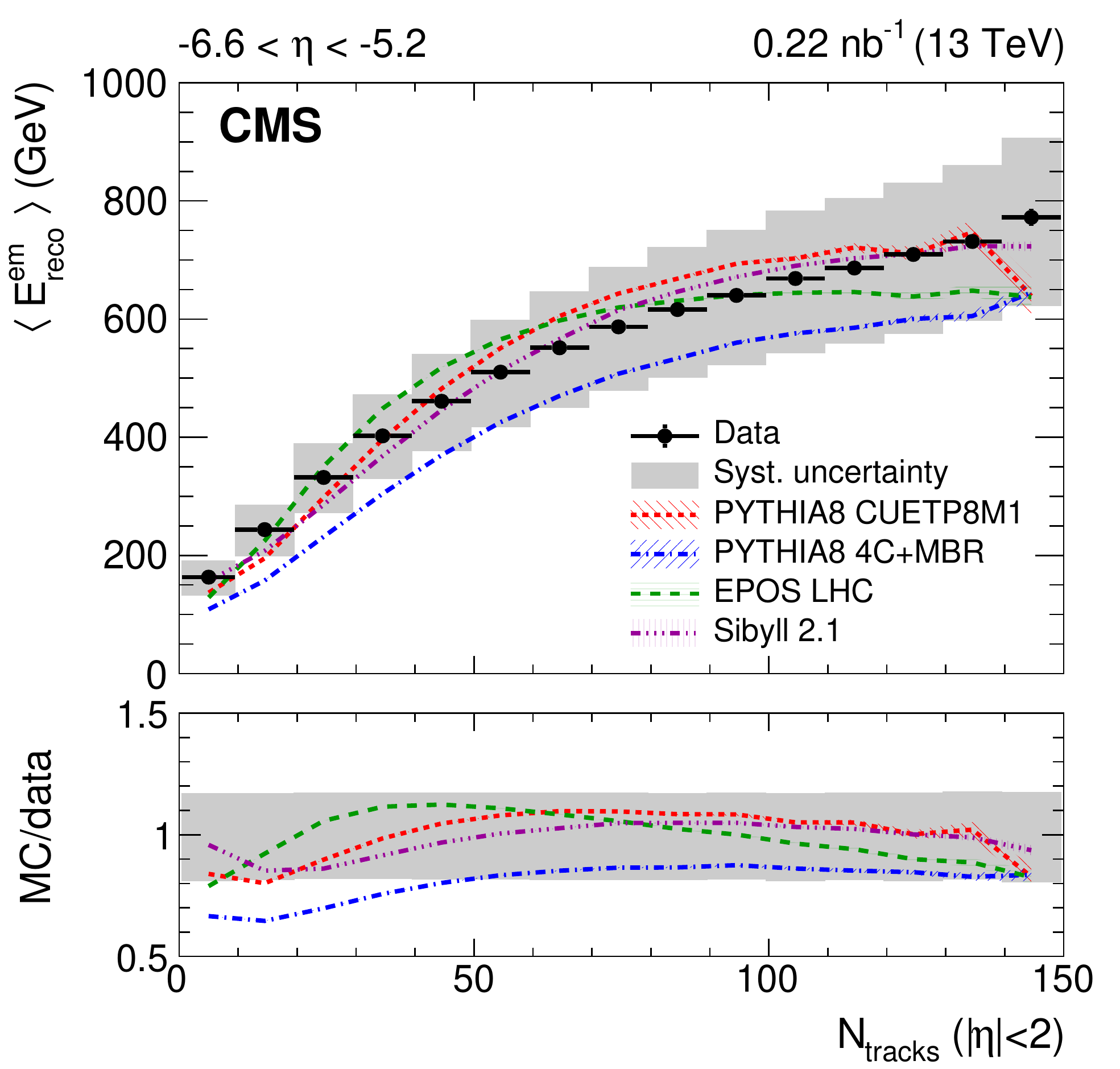}~
\includegraphics[width=0.45\linewidth]{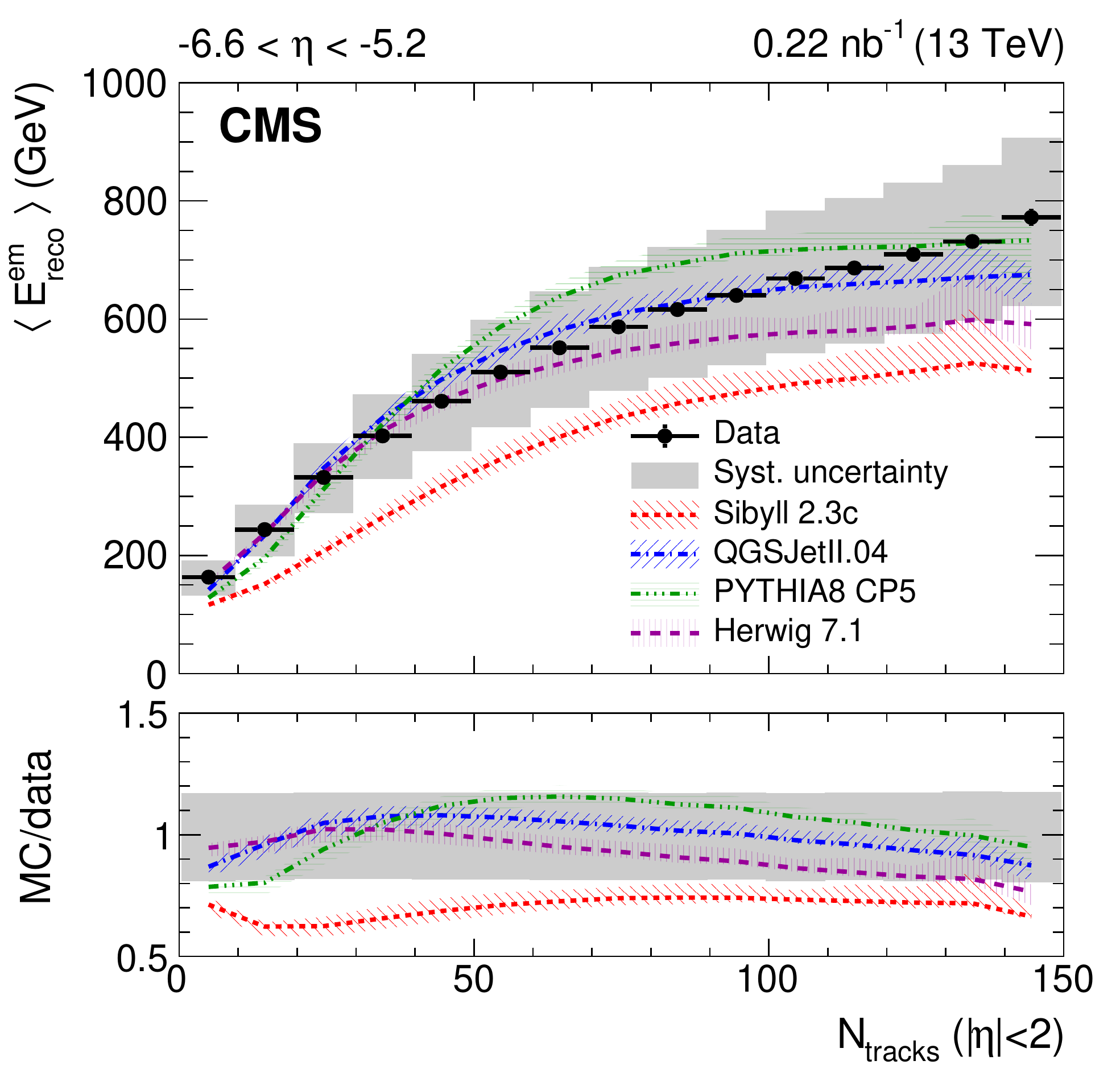}\\
\includegraphics[width=0.45\linewidth]{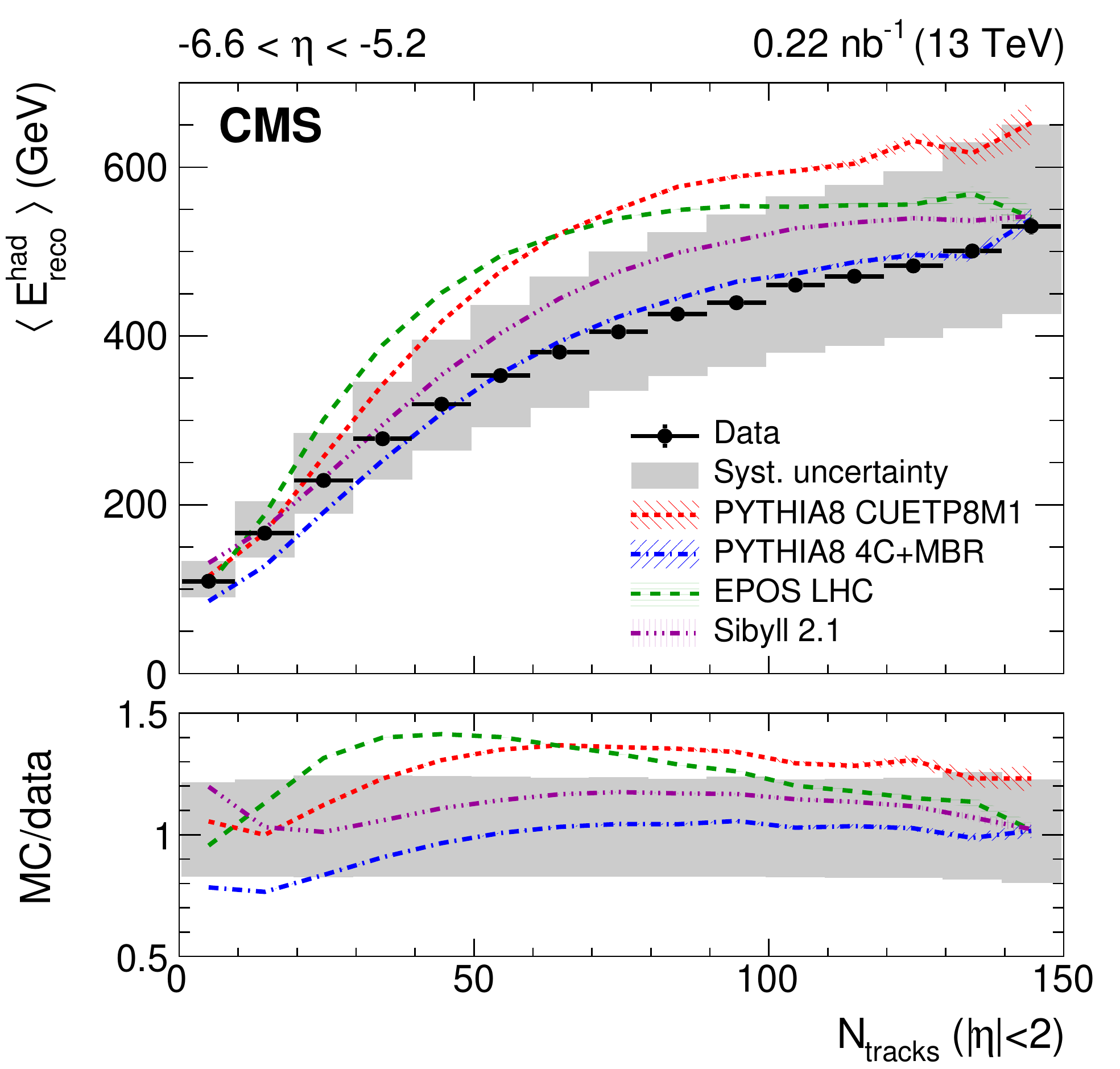}~
\includegraphics[width=0.45\linewidth]{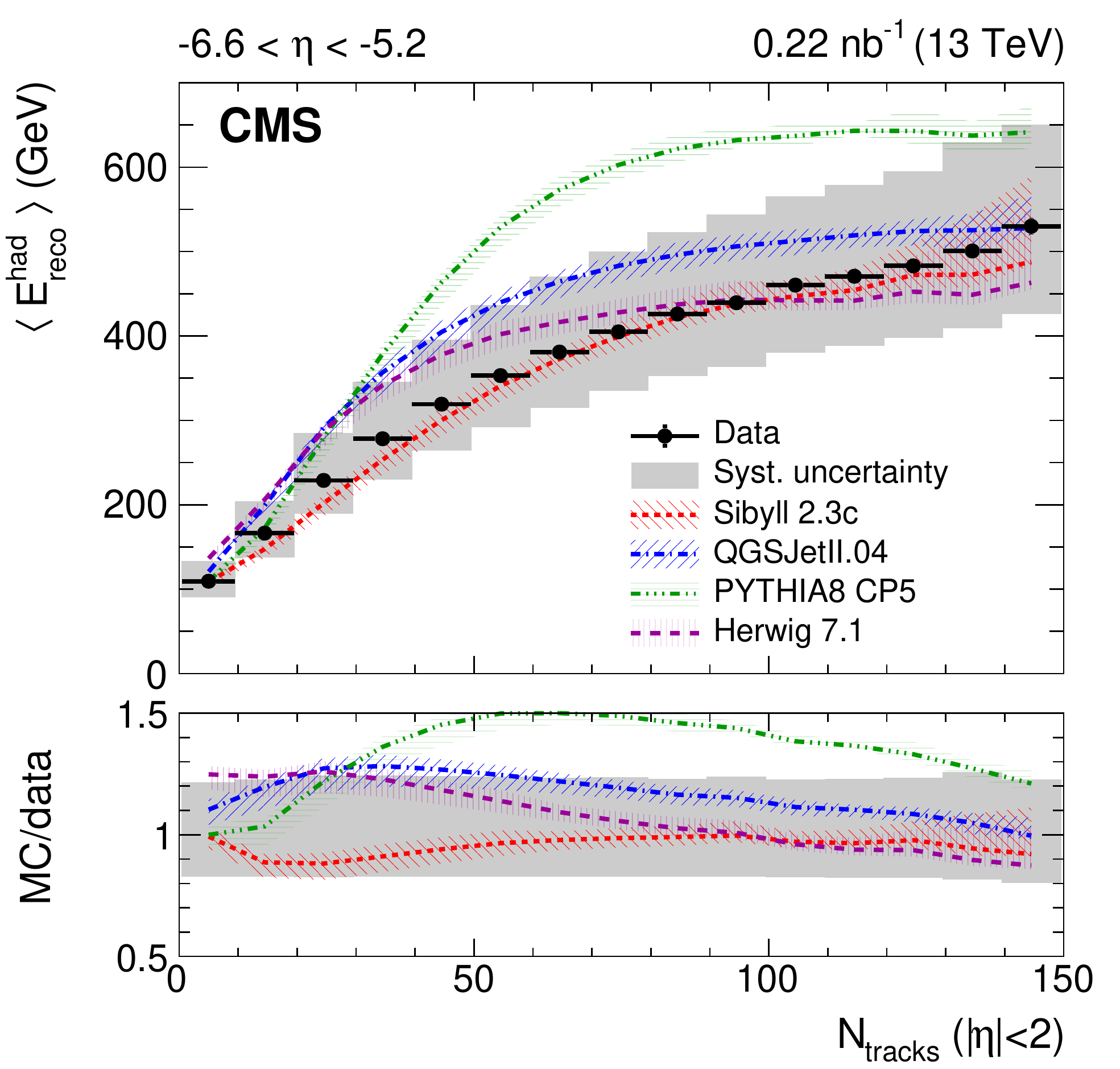}\\
\caption{Top panel: Average electromagnetic energy reconstructed in the CASTOR
calorimeter as a function of the number of reconstructed tracks for
$\abs{\eta}<2$.
Bottom panel: Average hadronic energy reconstructed in the CASTOR calorimeter
as a function of the number of reconstructed tracks for $\abs{\eta}<2$.
In all figures, the data are shown with black circles and the corresponding
systematic uncertainties with a gray band; horizontal bars are used to
indicate the bin width. The predictions of various event
generators are compared to the data, which are the same in both panels. The bands associated with the
model predictions illustrate the model uncertainty.}
\label{fig:MFlowDetector_emhad}
\end{figure*}

\begin{figure*}[ptb]
\centering
\includegraphics[width=0.45\textwidth]{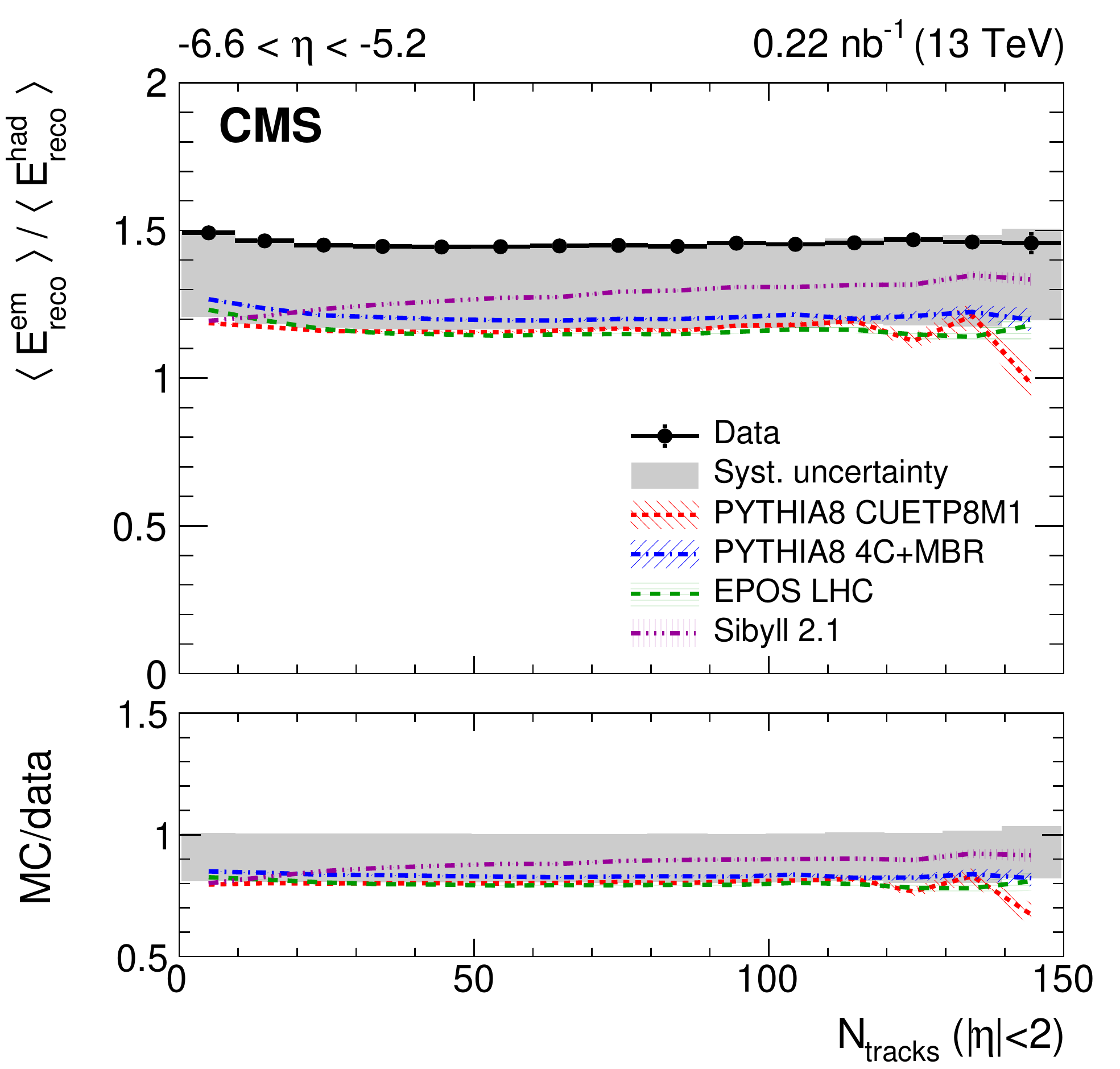}
\includegraphics[width=0.45\textwidth]{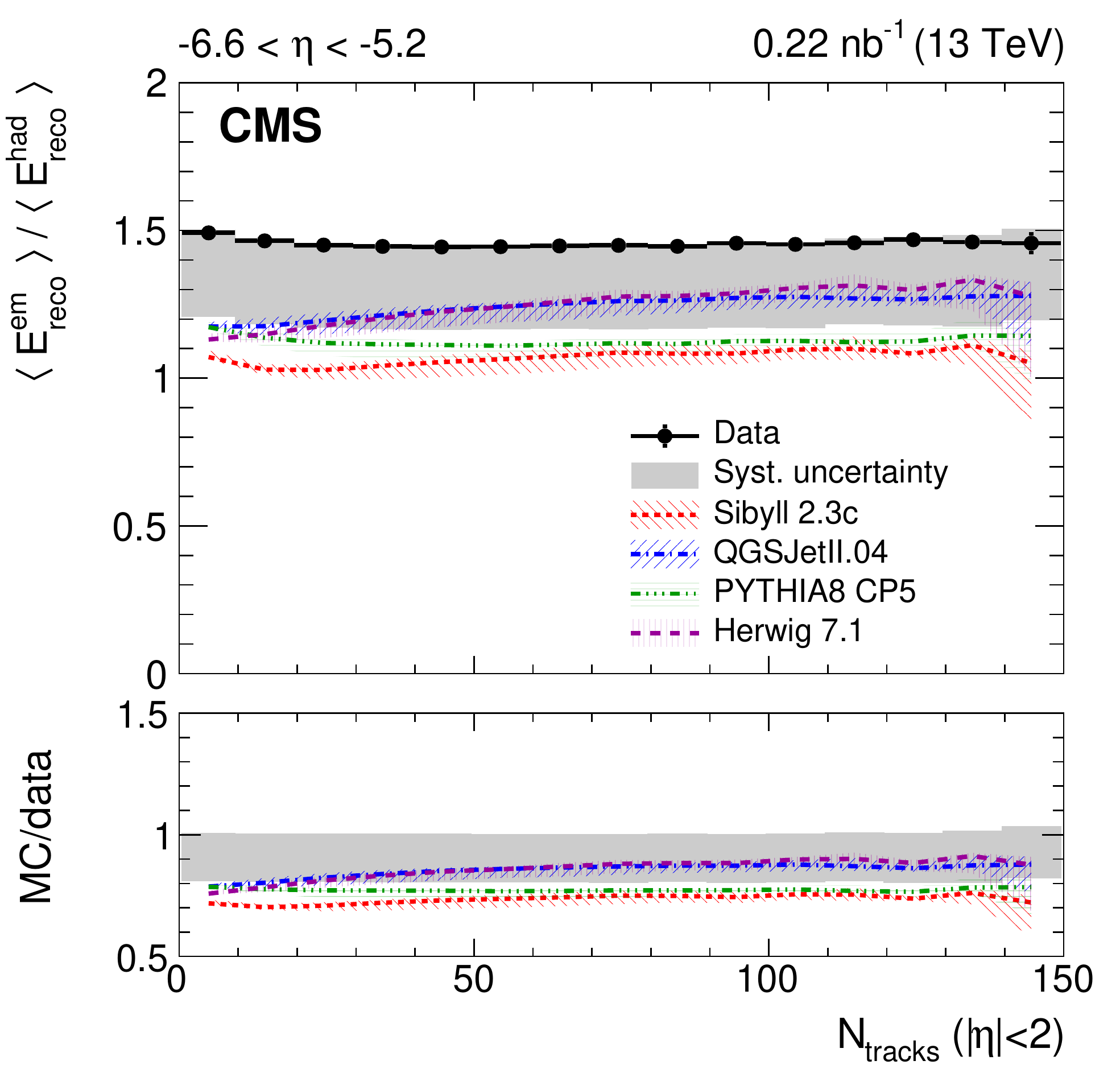}
\caption{Ratio of average electromagnetic and hadronic energies reconstructed
in the CASTOR calorimeter as a function of the number of reconstructed tracks
for $\abs{\eta}<2$. The data are shown with black circles and the
corresponding systematic uncertainties with a gray band; horizontal bars
are used to indicate the bin width. Predictions
of various  event generators are compared to the data, which are the same in both panels. The bands associated with the
model predictions illustrate the model uncertainty.}
\label{fig:MFlowDetector_ratio}
\end{figure*}

Various measurements of the average energy reconstructed in the region $-6.6<\eta<-5.2$
are presented as a function of the track
multiplicity for $\abs{\eta}<2$ in
Figs.~\ref{fig:MFlowDetector_all}--\ref{fig:MFlowDetector_ratio}. The
statistical uncertainties of the data are small and therefore not visible. The
systematic uncertainties are shown with a gray band.  The data are not
corrected for detector effects and are compared to the predictions of models commonly used to
describe hadron interactions at the LHC
and in high energy cosmic ray air showers. These models are grouped into two sets:

The first contains \PYTHIA8 tune CUETP8M1 and
tune \textsc{4C+MBR}, \epos{} and \sibyll~2.1. All these have a full
detector simulation. The error bands shown for these
models reflect only  the Monte Carlo statistical uncertainties. These
become visible especially in the last bin.

{\tolerance=550
The second set of models consists of \sibyll~2.3c, \qgsjetII.04, \PYTHIA8 tune CP5,
and \HERWIG~7.1. Predictions from these models are obtained using
the forward-folding method. The uncertainty bands shown for these
models also include the systematic uncertainties from the
forward-folding procedure discussed in the previous section.\par}

{\tolerance=950
The average total energy in CASTOR, shown in Fig.~\ref{fig:MFlowDetector_all}~(upper), increases with the track
multiplicity.  This feature is consistent with the general behaviour
of the underlying event measured at central rapidities~(see for
example Refs.~\cite{Aaboud:2017fwp,Sirunyan:2017vio}) and is
reproduced by all models.
The rise can be associated to an initial correlation of
central and forward event activity, which is damped by energy conservation
in the most violent collisions.
All models describe these data with at most minor discrepancies. This
implies that the model parameters for the underlying event determined
at central rapidities are valid also for the very forward data.
In detail, the energies predicted by \PYTHIA8~\textsc{4C+MBR} and
\sibyll~2.3c are slightly too low at small multiplicity. Conversely, at intermediate multiplicities,
\PYTHIA8~CP5 predicts average energies larger than those observed.
\par}

The systematic uncertainty in the data is dominated by the energy
scale uncertainty contribution, which is fully correlated between the
multiplicity bins. Therefore, the distributions can be normalised to
the first bin, so that, when comparing their shapes, the systematic
uncertainty is significantly smaller
(cf. Fig.~\ref{fig:MFlowDetector_all}, lower).
The rise is steep at low multiplicities and becomes more gradual at higher multiplicities.
All \PYTHIA8 tunes have very similar
shapes, inconsistent with that observed in the data. The disagreement
is strongest for \PYTHIA8~CP5, a tune optimised on underlying event
data at central rapidity. This tune uses parton distribution functions
at next-to-next-to-leading order and features a softer MPI cutoff
compared to \PYTHIA8 CUETP8M1 (see Ref.~\cite{GEN-17-001} for
details). The data therefore provide relevant information for future
generator improvements and tunes. The \epos{}, \qgsjetII.04,
and \HERWIG7.1 models predict saturation at multiplicities above 80, which is
not seen in the data.  Both versions of \sibyll{} provide predictions
in agreement with the data.

The individual electromagnetic and hadronic energy distributions are shown in
Figs.~\ref{fig:MFlowDetector_emhad}~(upper)
and \ref{fig:MFlowDetector_emhad}~(lower). All models, with the
exception of \sibyll~2.3c, describe the electromagnetic component
well. \PYTHIA8~\textsc{4C+MBR} slightly underestimates the
electromagnetic energy at low multiplicities.  Conversely, the other models
tend to overestimate the hadronic component.  Specifically these data can be very
relevant for improving the simulation of cosmic ray
induced extensive air showers, and specifically the modelling of the
production of neutral versus charged pions or other hadrons with longer lifetimes, since the energies
in the region $-6.6<\eta<-5.2$ are close to
those in the peak of the forward energy flow.

The data are also used to determine the ratio of the average
electromagnetic and hadronic energies
(Fig.~\ref{fig:MFlowDetector_ratio}). Here, the relative calibration
of the electromagnetic and hadronic sections is the main source of
uncertainty and results in a very asymmetric uncertainty band.  The
measured ratio is approximately constant over the whole multiplicity
range.  The ratio is sensitive to the details of hadronisation, and
discrepancies between models and data may reflect an inadequate
description of the hadron production mechanisms. String fragmentation,
remnant fragmentation, initial- or final-state radiation, the
effects of a possible very dense hydrodynamical phase, or the decay
of short-lived resonances may be relevant
to the understanding of the data.  The observed independence
of the measured ratio of
track multiplicity indicates that no dramatic change of the particle
production mechanism is observed at this very forward
pseudorapidity. All model predictions are lower than the data,
specifically those of the modern tunes \PYTHIA8~CP5 and \sibyll~2.3c,
whereas
\qgsjetII.04, \sibyll~2.1, and \HERWIG7.1 provide the best
description of the ratio.

\section{Summary and discussion}
\label{sec:summary}
The average energy per event in the pseudorapidity
region \CASTORAcceptance{} was measured as a function of the
observed central track multiplicity ($\abs{\eta}<2$) in proton-proton
collisions at a centre-of-mass energy of 13\TeV.  The data are recorded during
the first days of 13\TeV running with low beam intensities.
The measurement is presented in terms of the total energy as well as
its electromagnetic and hadronic components.  The very forward region
covered by the data contains the highest energy densities studied in
proton-proton collisions at the LHC so far. This makes the present
data relevant for improving the modelling of multiparticle production
in event generators of ultra-high energy cosmic ray air showers.

{\tolerance=450
The measured average total energy as a function of the track
multiplicity is described by all models reasonably well.
This demonstrates that the underlying event parameter
tunes determined at central rapidity can be safely extrapolated to the
very forward region within experimental uncertainties.
A shape analysis indicates, however, that there are
significant differences among the models and large deviations from the
data. The generator \sibyll~2.1 gives the best description of the measured
multiplicity dependence of the average total energy.
\par}

The data are also presented in terms of the average electromagnetic
and hadronic energies per event as a function of the central track
multiplicity. This is useful in the study of different particle production
mechanisms, since the former is primarily due to the decay of neutral
pions and the latter to the production of hadrons with longer lifetimes, mostly charged
pions. All
models give a good description of the electromagnetic
energy dependence on the multiplicity, with the exception
of \sibyll~2.3c. Conversely, the predictions for the hadronic energy
have a significantly larger spread compared to the electromagnetic
case.

The ratio between the electromagnetic and hadronic energies is also
presented. The data exhibit a larger fraction of electromagnetic
energy than the models, and disagree with the two most recent
model tunes, \ie \sibyll~2.3c and \PYTHIA8~CP5. Therefore, these models
cannot explain the muon deficit in ultra-high energy air shower
simulations since the data indicate that even more energy must be
channelled into the electromagnetic
part of the cascade and is thus lost for the generation of further
hadrons~\cite{Ulrich:2010rg}.

\begin{acknowledgments}
We congratulate our colleagues in the CERN accelerator departments for the excellent performance of the LHC and thank the technical and administrative staffs at CERN and at other CMS institutes for their contributions to the success of the CMS effort. In addition, we gratefully acknowledge the computing centres and personnel of the Worldwide LHC Computing Grid for delivering so effectively the computing infrastructure essential to our analyses. Finally, we acknowledge the enduring support for the construction and operation of the LHC and the CMS detector provided by the following funding agencies: BMBWF and FWF (Austria); FNRS and FWO (Belgium); CNPq, CAPES, FAPERJ, FAPERGS, and FAPESP (Brazil); MES (Bulgaria); CERN; CAS, MoST, and NSFC (China); COLCIENCIAS (Colombia); MSES and CSF (Croatia); RPF (Cyprus); SENESCYT (Ecuador); MoER, ERC IUT, PUT and ERDF (Estonia); Academy of Finland, MEC, and HIP (Finland); CEA and CNRS/IN2P3 (France); BMBF, DFG, and HGF (Germany); GSRT (Greece); NKFIA (Hungary); DAE and DST (India); IPM (Iran); SFI (Ireland); INFN (Italy); MSIP and NRF (Republic of Korea); MES (Latvia); LAS (Lithuania); MOE and UM (Malaysia); BUAP, CINVESTAV, CONACYT, LNS, SEP, and UASLP-FAI (Mexico); MOS (Montenegro); MBIE (New Zealand); PAEC (Pakistan); MSHE and NSC (Poland); FCT (Portugal); JINR (Dubna); MON, RosAtom, RAS, RFBR, and NRC KI (Russia); MESTD (Serbia); SEIDI, CPAN, PCTI, and FEDER (Spain); MOSTR (Sri Lanka); Swiss Funding Agencies (Switzerland); MST (Taipei); ThEPCenter, IPST, STAR, and NSTDA (Thailand); TUBITAK and TAEK (Turkey); NASU and SFFR (Ukraine); STFC (United Kingdom); DOE and NSF (USA).

\hyphenation{Rachada-pisek} Individuals have received support from the Marie-Curie programme and the European Research Council and Horizon 2020 Grant, contract Nos.\ 675440, 752730, and 765710 (European Union); the Leventis Foundation; the A.P.\ Sloan Foundation; the Alexander von Humboldt Foundation; the Belgian Federal Science Policy Office; the Fonds pour la Formation \`a la Recherche dans l'Industrie et dans l'Agriculture (FRIA-Belgium); the Agentschap voor Innovatie door Wetenschap en Technologie (IWT-Belgium); the F.R.S.-FNRS and FWO (Belgium) under the ``Excellence of Science -- EOS" -- be.h project n.\ 30820817; the Beijing Municipal Science \& Technology Commission, No. Z181100004218003; the Ministry of Education, Youth and Sports (MEYS) of the Czech Republic; the Lend\"ulet (``Momentum") Programme and the J\'anos Bolyai Research Scholarship of the Hungarian Academy of Sciences, the New National Excellence Program \'UNKP, the NKFIA research grants 123842, 123959, 124845, 124850, 125105, 128713, 128786, and 129058 (Hungary); the Council of Science and Industrial Research, India; the HOMING PLUS programme of the Foundation for Polish Science, cofinanced from European Union, Regional Development Fund, the Mobility Plus programme of the Ministry of Science and Higher Education, the National Science Center (Poland), contracts Harmonia 2014/14/M/ST2/00428, Opus 2014/13/B/ST2/02543, 2014/15/B/ST2/03998, and 2015/19/B/ST2/02861, Sonata-bis 2012/07/E/ST2/01406; the National Priorities Research Program by Qatar National Research Fund; the Ministry of Science and Education, grant no. 3.2989.2017 (Russia); the Programa Estatal de Fomento de la Investigaci{\'o}n Cient{\'i}fica y T{\'e}cnica de Excelencia Mar\'{\i}a de Maeztu, grant MDM-2015-0509 and the Programa Severo Ochoa del Principado de Asturias; the Thalis and Aristeia programmes cofinanced by EU-ESF and the Greek NSRF; the Rachadapisek Sompot Fund for Postdoctoral Fellowship, Chulalongkorn University and the Chulalongkorn Academic into Its 2nd Century Project Advancement Project (Thailand); the Welch Foundation, contract C-1845; and the Weston Havens Foundation (USA).
\end{acknowledgments}

\bibliography{auto_generated}

\cleardoublepage \appendix\section{The CMS Collaboration \label{app:collab}}\begin{sloppypar}\hyphenpenalty=5000\widowpenalty=500\clubpenalty=5000\input{FSQ-18-001-authorlist.tex}\end{sloppypar}
\end{document}

%% file: FSQ-18-001-authorlist.tex
\vskip\cmsinstskip
\textbf{Yerevan Physics Institute, Yerevan, Armenia}\\*[0pt]
A.M.~Sirunyan$^{\textrm{\dag}}$, A.~Tumasyan
\vskip\cmsinstskip
\textbf{Institut f\"{u}r Hochenergiephysik, Wien, Austria}\\*[0pt]
W.~Adam, F.~Ambrogi, T.~Bergauer, J.~Brandstetter, M.~Dragicevic, J.~Er\"{o}, A.~Escalante~Del~Valle, M.~Flechl, R.~Fr\"{u}hwirth\cmsAuthorMark{1}, M.~Jeitler\cmsAuthorMark{1}, N.~Krammer, I.~Kr\"{a}tschmer, D.~Liko, T.~Madlener, I.~Mikulec, N.~Rad, J.~Schieck\cmsAuthorMark{1}, R.~Sch\"{o}fbeck, M.~Spanring, D.~Spitzbart, W.~Waltenberger, J.~Wittmann, C.-E.~Wulz\cmsAuthorMark{1}, M.~Zarucki
% ...
\vskip\cmsinstskip
\dag: Deceased\\
1:  Also at Vienna University of Technology, Vienna, Austria\\
% ...
